\documentclass[12pt]{article}
\pdfoutput=1
\usepackage[colorlinks,linkcolor=blue,citecolor=blue,bookmarks,bookmarksnumbered]{hyperref}
\usepackage[scaled=0.85]{helvet}
\usepackage{amsmath,amssymb,accents,mathrsfs}
\usepackage{graphicx,color}
\usepackage{multirow}
\usepackage{amsmath, amssymb, mathtools, slashed, amsfonts, ulem}
\usepackage[margin=1in]{geometry}

\definecolor{Green}  {rgb}{0.10,0.70,0.10} 
\definecolor{Orange} {rgb}{1.00,0.50,0.15} 
\definecolor{Red}    {rgb}{0.90,0.00,0.12} 
\definecolor{Purple} {rgb}{0.50,0.25,0.55} 
\definecolor{Turque} {rgb}{0.00,0.65,0.85} 
\definecolor{Blue}   {rgb}{0.00,0.00,1.00} 
\definecolor{Magenta}{rgb}{1.00,0.00,1.00} 
\definecolor{Gold}   {rgb}{1.00,0.75,0.25} 
\definecolor{Seaweed}{rgb}{0.01,0.24,0.09} 
\definecolor{Brown}  {rgb}{0.43,0.26,0.32} 
\definecolor{grey1}  {rgb}{0.20,0.20,0.20} 
\definecolor{grey2}  {rgb}{0.40,0.40,0.40} 
\definecolor{grey3}  {rgb}{0.60,0.60,0.60} 
\definecolor{grey4}  {rgb}{0.80,0.80,0.80} 
\definecolor{grey5}  {rgb}{0.90,0.90,0.90} 
\def\C#1#2{{\ifcase#1\or
             \color{Green}\or \color{Orange}\or \color{Red}\or
              \color{Purple}\or \color{Turque}\or \color{Blue}\or
               \color{Magenta}\or \color{Gold}\or \color{Seaweed}\or
                \color{Brown}\or\color{grey1}\or\color{grey2}\or
                 \color{grey3}\else\color{grey4}\fi#2}}

\definecolor{Slate} {rgb}{0.00,0.45,0.55}

\def\rI{{\rm I}}
\def\rJ{{\rm J}}

\def\rL{{\rm L}}
\def\rR{{\rm R}}

\def\fracm#1#2{\hbox{\large{${\frac{{#1}}{{#2}}}$}}}

\def\be{\begin{equation}}
\def\ee{\end{equation}}
\newcommand{\bea}{\begin{eqnarray}}
\newcommand{\eea}{\end{eqnarray}}
\newcommand{\ena}{\end{eqnarray}}

\def\pp{{\mathchoice
          {
              \kern 1pt%
              \raise 1pt
              \vbox{\hrule width5pt height0.4pt depth0pt
                    \kern -2pt
                    \hbox{\kern 2.3pt
                          \vrule width0.4pt height6pt depth0pt
                          }
                    \kern -2pt
                    \hrule width5pt height0.4pt depth0pt}%
                    \kern 1pt
           }
            {
              \kern 1pt%
              \raise 1pt
              \vbox{\hrule width4.3pt height0.4pt depth0pt
                    \kern -1.8pt
                    \hbox{\kern 1.95pt
                          \vrule width0.4pt height5.4pt depth0pt
                          }
                    \kern -1.8pt
                    \hrule width4.3pt height0.4pt depth0pt}%
                    \kern 1pt
            }
            {
              \kern 0.5pt%
              \raise 1pt
              \vbox{\hrule width4.0pt height0.3pt depth0pt
                    \kern -1.9pt  
                    \hbox{\kern 1.85pt
                          \vrule width0.3pt height5.7pt depth0pt
                          }
                    \kern -1.9pt
                    \hrule width4.0pt height0.3pt depth0pt}%
                    \kern 0.5pt
            }
            {
              \kern 0.5pt%
              \raise 1pt
              \vbox{\hrule width3.6pt height0.3pt depth0pt
                    \kern -1.5pt
                    \hbox{\kern 1.65pt
                          \vrule width0.3pt height4.5pt depth0pt
                          }
                    \kern -1.5pt
                    \hrule width3.6pt height0.3pt depth0pt}%
                    \kern 0.5pt
            }
        }}

\def\mm{{\mathchoice
                       {
                             \kern 1pt
               \raise 1pt    \vbox{\hrule width5pt height0.4pt depth0pt
                                  \kern 2pt
                                  \hrule width5pt height0.4pt depth0pt}
                             \kern 1pt}
                       {
                            \kern 1pt
               \raise 1pt \vbox{\hrule width4.3pt height0.4pt depth0pt
                                  \kern 1.8pt
                                  \hrule width4.3pt height0.4pt depth0pt}
                             \kern 1pt}
                       {
                            \kern 0.5pt
               \raise 1pt
                            \vbox{\hrule width4.0pt height0.3pt depth0pt
                                  \kern 1.9pt
                                  \hrule width4.0pt height0.3pt depth0pt}
                            \kern 1pt}
                       {
                           \kern 0.5pt
             \raise 1pt  \vbox{\hrule width3.6pt height0.3pt depth0pt
                                  \kern 1.5pt
                                  \hrule width3.6pt height0.3pt depth0pt}
                           \kern 0.5pt}
                       }}

\def\ad{{\kern0.5pt
                   \alpha \kern-5.05pt \raise5.8pt\hbox{$\textstyle.$}\kern
0.5pt}}

\def\bd{{\kern0.5pt
                   \beta \kern-5.05pt \raise5.8pt\hbox{$\textstyle.$}\kern
0.5pt}}

\def\qd{{\kern0.5pt
                   q \kern-5.05pt \raise5.8pt\hbox{$\textstyle.$}\kern
0.5pt}}
\def\Dot#1{{\kern0.5pt
     {#1} \kern-5.05pt \raise5.8pt\hbox{$\textstyle.$}\kern
0.5pt}}

\catcode`@=11
\def\un#1{\relax\ifmmode\@@underline#1\else
        $\@@underline{\hbox{#1}}$\relax\fi}
\catcode`@=12

\def\g{\gamma}

\def\l{\lambda}
\def\m{\mu}

\def\x{\xi}
\def\z{\zeta}

\def\dslash{\not{\hbox{\kern-2pt $\partial$}}}
\def\Dslash{\not{\hbox{\kern-4pt $D$}}}
\def\pslash{\not{\hbox{\kern-2.3pt $p$}}}
 \newtoks\slashfraction
 \slashfraction={.13}
 \def\slash#1{\setbox0\hbox{$ #1 $}
 \setbox0\hbox to \the\slashfraction\wd0{\hss \box0}/\box0 }
 
\def\kcr{{\hbox{\ro \char'170}}}                
\def\ktl{{\hbox{\ro \char'170}}}        
\def\ktr{{\hbox{\ro \char'170}}}        
\def\kbl{{\hbox{\ro \char'170}}}        
\def\kbr{{\hbox{\ro \char'170}}}        

\def\plpl{\raise-2pt\hbox{$\raise3pt\hbox{$_+$}\hskip-6.67pt\raise0.0pt
\hbox{$^+$}\hskip 0.01pt$}}
\def\mimi{\raise-2pt\hbox{$\raise3pt\hbox{$_-$}\hskip-6.67pt\raise0.0pt
\hbox{$^-$}\hskip 0.01pt$}} 

\def\bo{{\raise.15ex\hbox{\large$\Box$}}}               
\def\pa{\partial}                                       
\def\TH{{\raise.2ex\hbox{$\displaystyle \bigodot$}\mskip-4.7mu \llap H \;}}
\def\face{{\raise.2ex\hbox{$\displaystyle \bigodot$}\mskip-2.2mu \llap {$\ddot
        \smile$}}}                                      

\def\dt#1{\on{\hbox{\bf .}}{#1}}                
\def\Dot#1{\dt{#1}}

\def\sb#1{{}_{#1}}                              
\def\Hat#1{\widehat{#1}}                        
\def\leftrightarrowfill{$\mathsurround=0pt \mathord\leftarrow \mkern-6mu
        \cleaders\hbox{$\mkern-2mu \mathord- \mkern-2mu$}\hfill
        \mkern-6mu \mathord\rightarrow$}
\def\dvec#1{\vbox{\ialign{##\crcr
        \leftrightarrowfill\crcr\noalign{\kern-1pt\nointerlineskip}
        $\hfil\displaystyle{#1}\hfil$\crcr}}}           
\def\dt#1{{\buildrel {\hbox{\LARGE .}} \over {#1}}}     

\def\fracm#1#2{\hbox{\large{${\frac{{#1}}{{#2}}}$}}}
\def\sfrac#1#2{{\vphantom1\smash{\lower.5ex\hbox{\small$#1$}}\over
        \vphantom1\smash{\raise.4ex\hbox{\small$#2$}}}} 
\def\bfrac#1#2{{\vphantom1\smash{\lower.5ex\hbox{$#1$}}\over
        \vphantom1\smash{\raise.3ex\hbox{$#2$}}}}       
\def\afrac#1#2{{\vphantom1\smash{\lower.5ex\hbox{$#1$}}\over#2}}    

\def\pa{\partial}      
\let\bm\relax
\newcommand{\bm}[1]{{\boldsymbol{#1}}}

\def\ad{{\dot{\alpha}}}
\def\bd{{\dot{\beta}}}

 \font\rOpe=cmsy10                        
 \def\ktl{{\hbox{\rOpe\char'170}}}        
 \def\kbl{{\hbox{\rOpe\char'170}}}        
 \def\kcr{{\reflectbox{\rOpe\char'170}}}        
 \def\ktr{{\reflectbox{\rOpe\char'170}}}        
 \def\kbr{{\reflectbox{\rOpe\char'170}}}        
 \def\Border{\vbox{\hsize0pt
        \setlength{\unitlength}{1mm}
        \newcount\xco
        \newcount\yco
        \xco=-21
        \yco=12
        \begin{picture}(0,0)(-7.5,0)
        \put(\xco,\yco){$\ktl$}
        \advance\yco by-1
        {\loop
        \put(\xco,\yco){$\kcr$}
        \advance\yco by-2
        \ifnum\yco>-240
        \repeat
        \put(\xco,\yco){$\kbl$}}
        \xco=170
        \yco=12
        \put(\xco,\yco){$\ktr$}
        \advance\yco by-1
        {\loop
        \put(\xco,\yco){$\kcr$}
        \advance\yco by-2
        \ifnum\yco>-240
        \repeat
        \put(\xco,\yco){$\kbr$}}
        \put(-19.5,13){\scalebox{.6065}{%
         University of Maryland Center for String and Particle  Theory \&\ Physics Department%
        |University of Maryland Center for String and Particle  Theory \&\ Physics Department}}
        \put(-19.5,-241.5){\scalebox{.5835}{%
         ****University of Maryland * Center for String and
         Particle  Theory* Physics Department****University of Maryland *Center
        for String and Particle  Theory* Physics Department}}
        \end{picture}
        \par\vskip-8mm}}
\definecolor{UMred}{rgb}{.9,.05,.2}
\definecolor{HUblue}{rgb}{.0,.3,.7}

\definecolor{Red}    {rgb}{0.90,0.00,0.12} 
\definecolor{Blue}   {rgb}{0.00,0.00,1.00} 
\definecolor{Green}  {rgb}{0.10,0.70,0.10} 
\definecolor{Turque} {rgb}{0.00,0.65,0.85} 
\definecolor{Orange} {rgb}{1.00,0.50,0.15} 
\definecolor{Magenta}{rgb}{1.00,0.00,1.00} 
\definecolor{Gold}   {rgb}{1.00,0.75,0.25} 
\definecolor{Seaweed}{rgb}{0.01,0.24,0.09} 
\definecolor{Purple} {rgb}{0.50,0.25,0.55} 
\definecolor{Brown}  {rgb}{0.43,0.26,0.32} 
\definecolor{grey1}  {rgb}{0.20,0.20,0.20} 
\definecolor{grey2}  {rgb}{0.40,0.40,0.40} 
\definecolor{grey3}  {rgb}{0.60,0.60,0.60} 
\definecolor{grey4}  {rgb}{0.80,0.80,0.80} 
\definecolor{grey5}  {rgb}{0.90,0.90,0.90} 
\def\C#1#2{{\ifcase#1\or
             \color{Red}\or \color{Green}\or \color{Blue}\or\
              \color{Turque}\or \color{Orange}\or \color{Magenta}\or 
               \color{Gold}\or \color{Seaweed}\or \color{Purple}\or
                \color{Brown}\or\color{grey1}\or\color{grey2}\or
                 \color{grey3}\else\color{grey4}\fi#2}}

\definecolor{Slate} {rgb}{0.00,0.45,0.55}

\newdimen\parshift\parshift=\parindent
\catcode`@=11
 \long\def\@footnotetext#1{\insert\footins{\reset@font\footnotesize
           \interlinepenalty\interfootnotelinepenalty\splittopskip%
            \footnotesep\splitmaxdepth\dp\strutbox\floatingpenalty\@MM%
             \hsize\columnwidth\addtolength{\hsize}{-2\parindent}
              \@parboxrestore\protected@edef\@currentlabel%
              {\csname p@footnote\endcsname\@thefnmark}%
                \color@begingroup%
                 \@makefntext{\rule\z@\footnotesep\ignorespaces#1%
                  \@finalstrut\strutbox}%
                \color@endgroup}}
 \long\def\@makefntext#1{\hglue\parshift%
           \vbox{\noindent\baselineskip=11pt plus.5pt minus.5pt\hb@xt@0em{\hss\@makefnmark\kern1pt}#1}}
\catcode`@=12

\newskip\humongous \humongous=0pt plus 1000pt minus 1000pt

\newif\ifdtup

\makeatletter
\def\section{\@startsection{section}{1}{\z@}
        {3ex plus-1ex minus-.2ex}{1pt plus1pt}{\large\sf\bfseries\boldmath}}
\def\subsection{\@startsection{subsection}{2}{\z@}
         {1.5ex plus-1ex minus-.2ex}{0.01pt plus1pt}{\sf\slshape}}
\def\subsubsection{\@startsection{subsubsection}{3}{\z@}
          {1.5ex plus-1ex minus-.2ex}{0.01pt plus0.2pt}{\sf\boldmath}}
\def\paragraph{\@startsection{paragraph}{4}{\z@}
           {.75ex \@plus.5ex \@minus.2ex}{-2mm}{\sf\bfseries\boldmath}}
\makeatother

\allowdisplaybreaks
\numberwithin{figure}{section}

\usepackage{tikz}
\usepackage[vcentermath,enableskew]{youngtab}

\definecolor{Hey}{rgb}{.9,.05,.4}
\definecolor{orange}{rgb}{1,.5,0}
\definecolor{plum}{rgb}{.4,0,.6}
\definecolor{R}{rgb}{1,0,0}
\definecolor{G}{rgb}{0,1,0}
\definecolor{B}{rgb}{0,0,1}
\usepackage{lipsum}
\usepackage{listings}
\definecolor{MyDarkGreen}{rgb}{0.0,0.4,0.0} 
\newcommand{\LP}{\left(}
\newcommand{\RP}{\right)}

\newcommand{\R}{\mathbb{R}}

\newcommand{\B}[1]{\mathbf{#1}}

\def\rI{{\rm I}}
\def\rJ{{\rm J}}

\def\rL{{\rm L}}

\begin{document}

\thispagestyle{empty}
\noindent{\small
\hfill{$~~$}  \\ 
{}
}
\begin{center}
{\large \bf
Supersymmetry and Representation Theory in Low Dimensions 
}   \\   [8mm]
{\large {
Mathew Calkins\footnote{mathewpcalkins@google.com}${}^{a,b}$, 
S.\ James Gates, Jr.\footnote{sylvester$_-$gates@brown.edu}${}^{c, d}$, and
Caroline Klivans\footnote{Caroline$_-$Klivans@brown.edu}${}^{c,e,f}$
}}
\\*[6mm]

\emph{
\centering
$^{a}$Google Search,
\\[1pt]
111 8th Ave, 
 New York, NY 10011, USA, 
\\[10pt]
$^{b}$Courant Institute of Mathematical Sciences, New York University,
\\[1pt]
251 Mercer Street, New York, NY 10012, USA
\\[10pt]
$^{c}$Brown Theoretical Physics Center,
\\[1pt]
Box S, 340 Brook Street, Barus Hall,
Providence, RI 02912, USA
\\[10pt]
$^{d}$Department of Physics, Brown University,
\\[1pt]
Box 1843, 182 Hope Street, Barus \& Holley,
Providence, RI 02912, USA 
\\[10pt]
$^{e}$Division of Applied Mathematics, Brown University,
\\[1pt]
182 George Street,
Providence, RI 02906, USA
\\[10pt]
and
\\[10pt]
$^{f}$Institute for Computational \& Experimental Research in Mathematics, Brown University,
\\[1pt]
121 South Main Street
Providence, RI 02903, USA
}
 \\*[10mm]
{ ABSTRACT}\\[5mm]
\parbox{142mm}{\parindent=2pc\indent\baselineskip=14pt plus1pt
Beginning from a discussion of the known most fundamental dynamical structures of the 
Standard Model of physics, extended into the realms of mathematics and theory
by the concept of ``supersymmetry'' or ``SUSY,'' an introduction to efforts to develop
a complete representation theory is given.  Techniques drawing from graph theory,
coding theory, Coxeter Groups, Riemann surfaces, and computational approaches to the study of 
algebraic varieties are briefly highlighted as pathways for future exploration and
progress.
}
\end{center}
\vfill
\vfill
\clearpage

%

\newpage

\section{Supersymmetry and the Standard Model}
\indent
\
\\
As far as  experiments in elementary particle physics reveal, the basic constituents of matter and interactions (i.e. forces
excluding gravity) in our universe can be summarized in the list of particles shown in the table indicated in Fig.\  (1.1).

\begin{figure}[htp!]
\centering
\includegraphics[width=0.85\textwidth]{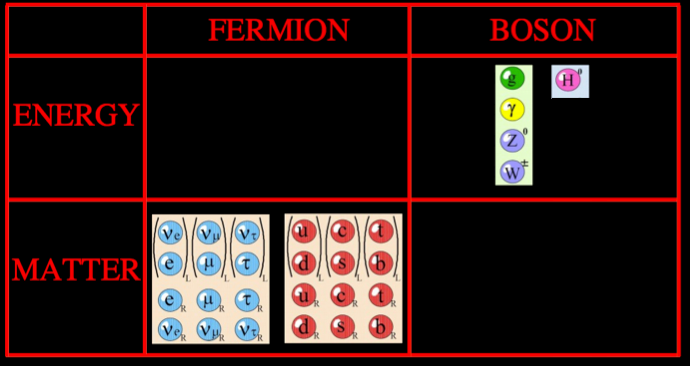}
\caption{Diagrammatic Representation of the `Standard Model' Particles }
\label{Fig:SM}
\end{figure}
\begin{figure}[htp!]
\centering
\includegraphics[width=0.85\textwidth]{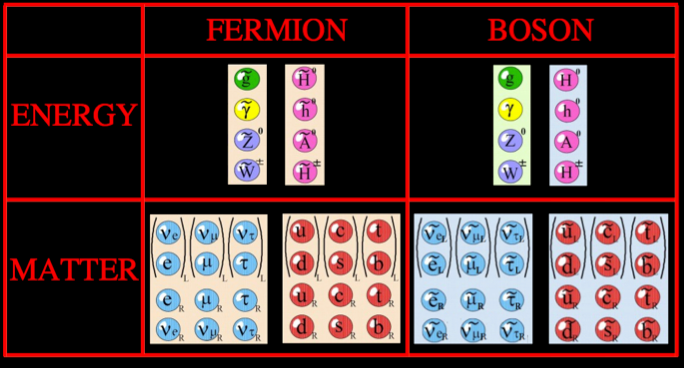}
\caption{Diagrammatic Representation of the `Minimal SUSY Standard Model' Particles }
\label{Fig:mSSM}
\end{figure}

\vskip0.01in
\
\\
\
The \textit{matter particles} consist of \textit{leptons} and \textit{quarks} shown in the lower left-hand quadrant. The most familiar member of the lepton family, shown within a light blue sphere, is the \textit{electron} indicated by the letter ``e.''   The \textit{up} and \textit{down quarks}, indicated by the letters ``u'' and ``d'' surrounded by red spheres, are the most familiar of the quark family. They are also the principal constituents of protons and neutrons.\\
\\
\
The \textit{gauge} or \textit{force-carrying particles} appear in the upper right-hand quadrant.  Each of the fundamental forces in nature is associated with one or more of these particles species.  The strong nuclear force is associated with eight \textit{gluons} indicated by the letter ``g'' within a green sphere.  The electromagnetic force is associated with the \textit{photon} indicated by the Greek letter ``$\g$'' within a yellow sphere.  Finally, for the weak nuclear force the \textit{Higgs boson} indicated by the letter ``H'' within a purple sphere, the \textit{Z-boson} indicated by the letter ``Z'' within a dark blue sphere, and two charged \textit{W-bosons} indicated by the symbols ``$W^+$''  and ``$W^-$'' also within dark blue spheres
are the associated force-carrying particle species.  The vertical heights of the quadrants of the elementary particles in Figure~(\ref{Fig:SM}) thus corresponds to whether each particle is a carrier of a fundamental force 
(upper level) or is a matter particle (lower level) subject to such a force. How then is the left-right arrangement of the particles significant?\\
\\
\
One of the great discoveries of quantum mechanics is elementary particles (with the exception of the Higgs boson) all possess \textit{spin}, a sort of intrinsic angular momentum  which augments the particle's more familiar orbital angular momentum.   The rate of spin is constrained to take the form $n \,[h / (4 \pi)]$, where $n$ is any nonnegative integer and $h$ is  ``Planck's Constant," a constant of nature empirically known to be roughly $6.626 \times 10^{-34}$ Joules $\times$ seconds. If the integer $n$ associated with a specific particle is even\footnote{If we include the value $n$ = 0, this covers the case of the Higgs Boson.} the particle is called a `boson,' if the 
integer is odd then the particle is called a `fermion.'  Thus, the left-right dichotomy shown in Figure~(\ref{Fig:SM}) 
is due to grouping of all fermions to the left half and all bosons in right half of the diagram.\\
\\
\
As technology advances, additional elementary particles continue to be discovered, most recently with the 2012 initial observation of the Higgs boson. It is likely this trend will continue. One possibility for future discovery is indicated in Figure~(\ref{Fig:mSSM}),
\vskip0.01in
\noindent
where a panoply of hypothetical new elementary particle species called \textit{superparticles}  (or more rarely \textit{sparticles}) are shown. An easily noticeable feature of Figure~(\ref{Fig:mSSM}), absent in Figure~(\ref{Fig:SM}), is its 
left-right symmetry.  This visual property is a reflection of a mathematical property called \textit{supersymmetry} \cite{allsusy,colemanmandula} or \textit{SUSY}. More formally, the possible states of a physics theory are elements of a Hilbert space.  This includes the particles
also described by such elements.   A theory is said to possess supersymmetry if there exist linear maps $Q: \mathcal{H} \to \mathcal{H}$ (sometimes called \textit{supercharges}) which take bosonic states to fermionic states and vice versa in a way that satisfies a certain algebra \cite{allsusy}.  We will treat these ideas more carefully in a later section.\\
\\
\
Today's experimentally confirmed particle theories do not feature supersymmetry.  Supersymmetry generally requires the addition of new particle species as seen in Figure~(\ref{Fig:mSSM}), so that the space of physical states and particles 
is closed under the action of each supercharge $Q$. However, there are many compelling reasons to suspect that  supersymmetry (such as consistency of a quantum theory of gravitation), may be a feature of nature.  Moreover by studying these speculative supersymmetric theories physicists have already extended knowledge about experimentally confirmed ones with greater precision.

\section{Formal Aspects of Supersymmetry}

\subsection{Supersymmetry Field Theory}
\
\\
Supersymmetric quantum field theory is of interest to physicists for its role in theories of particle physics beyond the Standard Model \cite{weinberg1,weinberg3,wessbagger} and of quantum gravity \cite{bbs}. It has also borne fruit in pure mathematics, see e.g., \cite{susymorse}. The full story of SUSY QFT is complex and beyond the scope of this work, not the
least because of the complexities of QFT in general, but the minimal version of the tale is as follows.\\
\\
\
Essentially all models of the physical universe consider the collection of all events, that is, ordered pairs $(ct, \B{x}) \equiv (x^0, x^1, x^2, x^3)$ of times and locations, as a smooth manifold aptly called \textit{spacetime}\footnote{Here $c$ is the speed of light measured to be to be roughly $2.99 \times 10^{8}$ meters/seconds}${}^,$
\footnote{One of they key insights Einstein's theory of relativity is that spacetime can be modeled as a smooth pseudo-Riemannian manifold, whence the gravitational field is described by the corresponding metric connection} In conventional QFT, we consider spacetime as it is modeled in the framework of \textit{special relativity}: a real 4-dimensional affine space equipped with an inner product which, by judicious choice of coordinate system, can be put globally into the form 
\begin{equation}
\eta(A, B) := A^0 B^0 - A^1B^1 - A^2B^2 - A^3B^3 ~~~, 
\label{Mnk}
\end{equation}
called the \textit{Minkowski metric} for two such ordered pairs $(A^0, A^1, A^2, A^3)$ and $(B^0, B^1, B^2, B^3)$.

To make a long and subtle story very short, it turns out that the inner product  is the core concept behind inertial reference frames: like any metric on any manifold, the numerical components of $\eta$ at any point in spacetime depend on the choice of local chart. Inertial reference frames are precisely those coordinate systems on spacetime in which the metric takes the above form at every point in the tangent space.  Spacetime coordinate changes such as turning your head, hopping onto a steadily moving train, or adjusting your watch for Daylight Savings Time - are  exactly those coordinate changes which preserve the form of the Minkowski metric, and which maintain the handedness of space and the direction of increasing time.  Such changes are known as \textit{Poincar\'{e}} transformations. \\

In order to reconcile these principles of special relativity with those of quantum mechanics, we consider the aforementioned Hilbert space acted upon by the \textit{Poincar\'{e} group} of metric-preserving transformations, or more precisely by its corresponding algebra.   In order to demonstrate the existence of representation of the Poincar\'{e} Lie algebra, a collection of operators $(P_{\mu})$ and $\LP M_{\mu \nu} \RP$, as well as their realization acting of the space of fields is introduced.

The physical states in the theory transform linearly in prescribed ways under the action of these generators, so that the classification of physical theories (before consideration of interactions) reduces to the classification of representations of the Poincar\'{e} algebra. This program is worked out in detail in \cite{weinberg1} and \cite{folland} in language amenable to physicists and mathematicians, respectively. From this point of view, a physical theory is said to possess supersymmetry if there exist operators $\LP Q_a^A \RP$\footnote{The index $a$ takes on values 1, $\dots$, 4 and describe a section of a spin bundle of the Clifford algebra $CL(1,3)$ while the index $A$ takes on values 1, $\dots$, $\cal N$ where $\cal N$ is any nonvanishing integer.} in the spin representation of the Poincar\'{e} algebra which together with $\LP P_{\mu} \RP $ and $\LP M_{\mu \nu} \RP$ satisfy the SUSY algebra given in \cite{allsusy}.

\subsection{Supersymmetric mechanics}
\
\\
The story simplifies significantly if  we project down to a spacetime with one temporal coordinate and no spatial coordinates (\textit{i.e.}, to the case of supersymmetric quantum mechanics).The Poincar\'{e} group of coordinate transformations then contains only translation along the time axis. This toy model features a collection of time-dependent bosonic fields\footnote{By ``field" we mean roughly a smooth distribution over the time axis $\R$ valued in the ring of linear maps $\mathcal{H} \to \mathcal{H}$. Since the exact identity of these objects will play a surprisingly small role in the present treatment, the reader is welcome to imagine them more simply as smooth, integrable maps $\R \to \R$.} $\phi_i(t)$ and fermionic fields $\psi_i(t)$ ($i = 1, \ldots, d$), which are respectively even and odd in the sense that products of fields satisfy 
\begin{align}
\mbox{bosonic} \times \mbox{bosonic} & = \mbox{bosonic} ~~~\,~~,  \nonumber \\
\mbox{bosonic} \times \mbox{fermionic} & = \mbox{fermionic} ~~~,  \nonumber \\
\mbox{fermionic} \times \mbox{fermionic} & = \mbox{bosonic} ~~~\,~~.
\end{align}
The collection is known as a \textit{supermultiplet} and the various fields its \textit{component fields}.\\
\\
\
The component fields (and by extensions, derivatives and linear combinations thereof) are subject to the action of a collection of linear operators $Q_\rI \ (\rI = 1, \ldots, \mathcal{N}$) called \textit{supercharges} which are odd in the sense of mapping bosonic fields to fermionic fields and vice versa. The appropriately reduced SUSY algebra then demands only that the action of the supercharges on the component fields ought to satisfy relations involving $\pa_0$ (the infinitesimal
generator of temporal translations) as shown below,
\begin{equation}
\{  \, Q_\rI ~,~ Q_\rJ  \, \} ~=~ i \, 2\,   \delta{}_{\rI \, \rJ} \,  \pa{}_{0}  ~~~, 
\end{equation}
and \begin{equation}
\left[ \, Q_\rI ~,~   \pa{}_{0} \, \right] ~=~ 0 ~~~.
\end{equation}
If these relations are valid independent of any differential equations imposed on the fields, the supermultiplet together with the charges acting on it is said to be an \textit{off-shell representation of $N$-extended 1-dimensional supersymmetry}. Since these relations constrain the ways the supercharges compose and add amongst themselves but say essentially nothing about the supermultiplet itself, we sometimes abuse vocabulary and refer to the supercharges alone as being such a representation.
\\
\\
\
These 1-dimensional toy models offer a fertile ground for studying the general structure of supersymmetric representations, not the least because (as we shall discuss shortly) they allow the problem to be recast into combinatorial and graph-theoretic terms. This richness makes it possible to study more physically realistic higher-dimensional models of supersymmetry by projecting down to 1-dimensional models.  Formally, this 
\textit{dimensional reduction} starts with a system of supercharges acting on time- and space-dependent fields $\phi_i (t, \B{x}), \psi_j(t,\B{x})$ and modifies their actions by formally setting all derivatives with respect to the spatial coordinates to zero, leaving only derivatives (if any) with respect to time.

\subsection{The Garden Algebra}
\
\\
In studying off-shell representations of $\mathcal{N}$-extended 1-dimensional supersymmetry, it is frequently useful to transform the component fields by invertible linear combinations 
\be{
\phi_i \to \sum_{j = 1}^d a_{ij} \phi_j ~~~, \ \ \ \psi_i \to \sum_{j = 1}^d b_{ij} \psi_j \ \ \ (a_{ij}, c_{ij} \in   {\mathbb{C}}) 
~~~~~~,
}\ee
or by differentiation
\begin{equation}
\phi_i \to \LP \pa{}_{0} \RP^{c_i} \phi_i ~~~, \ \ \ \psi_i \to \LP \pa{}_{0} \RP^{d_i} \psi_i \ \ \ \LP c_i, d_i \in \{0, 1\}\RP
~~~,
\end{equation}
both of which preserve satisfaction of the algebraic relations discussed above. Subject to some physically reasonable conditions discussed more thoroughly in \cite{generaloffshell}, it is generally possible to perform a sequence of such transformations so that the actions of the supercharges take the off-diagonal form 
\begin{equation}
Q_\rI (\phi_i) = \sum_{j = 1}^d (\rL_\rI)_{ij} \psi_j ~~~,~~~ \ \ \ Q_\rI (\psi_i) = i \sum_{j = 1}^d (\rR_\rI)_{ij} \pa{}_{0} \phi_j
~~~.
\end{equation}
or equivalently 
\begin{equation}
\begin{bmatrix}
\phi_1 \\
\vdots \\
\phi_d \\
\psi_1 \\
\vdots \\
\psi_d
\end{bmatrix} \xmapsto{Q_\rI} \begin{bmatrix}
0 & \rL_\rI \\
i\, \rR_\rI \circ \pa{}_{0} & 0 \\
\end{bmatrix}
\begin{bmatrix}
\phi_1 \\
\vdots \\
\phi_d \\
\psi_1 \\
\vdots \\
\psi_d
\end{bmatrix} ~~~.
\end{equation}
for some collection of matrices ${\bm {\rL}}_\rI, {\bm {\rR}}_\rI \in Gl(n,\R)$. In that case the condition $\{Q_\rI, Q_\rJ\} = 2 i \delta_{IJ} \pa{}_{0}$ reduces to the so-called \textit{Garden Algebra} 
\begin{align}
\label{gardenalgebra}
\begin{split}
{\bm {\rL}}_\rI \, {\bm {\rR}}_\rJ ~+~ {\bm {\rL}}_\rJ \, {\bm { \rR}}_\rI & ~=~ 2\,  \delta_{\rm {IJ}} \, \mathbf{1} ~~~, \\
{\bm {\rR}}_\rI \, {\bm {\rL}}_\rJ ~+~ {\bm {\rR}}_\rJ \, {\bm { \rL}}_\rI & ~=~ 2\,  \delta_{\rm {IJ}} \, \mathbf{1} ~~~, 
\end{split}
\end{align}
while the commutator relation $[Q_\rI \,,\, \pa{}_{0}] = 0$ holds automatically. Thus any collection of square matrices satisfying Eq. (\ref{gardenalgebra}) furnishes a representation of off-shell $(\mathcal{N},d)$ supersymmetry. The matrices ${\bm {\rL}}_\rI$ and ${\bm {\rR}}_\rI$ (with the index $\rI$ taking values of 1, $\dots$, $\mathcal{N}$) \cite{spinning1,spinning2} can also be used to construct $2d \times 2d$ matrices ${\Hat {\bm \g}}{}_\rI$ \cite{ENUF} as
\be{
{\Hat {\bm \g}}{}_\rI ~=~ \left[\begin{array}{cc}
~0 & ~~  {\bm {\rL}}_\rI  \\
{}~&~\\
~ {\bm {\rR}}_\rI & ~~ 0 \\
\end{array}\right]  ~~~,
} \label{CLFF} \ee
which form a Clifford Algebra, with respect to the Euclidean metric define by the Kronecker delta symbol
$ \delta_{\rm {IJ}}$
\begin{equation}
\left\{{\Hat {\bm \g}}{}_\rI, {\Hat {\bm \g}}{}_\rJ \right\} = 2 \delta_{\rm {IJ}} ~~~.
\label{CLFF2}
\end{equation}

\section{The Adinkra Program - Supersymmetry from Decorated Graphs}
\
\\
Now we proceed with a pair of worked examples. We shall find that they admit a concise description not only in terms of 
matrices, but in terms of certain decorated graphs. We begin with a note on a historical accident of convention.

\subsection{Charges and Derivatives}
\
\\
Thus far we have spoken of supersymmetric representations entirely in terms of the supercharges $Q_\rI$. It turns out that supersymmetry can be reformulated into the geometric language of \textit{superspace} \cite{wessbagger}, a sort of extension of spacetime to include fermionic coordinates in addition to the usual real bosonic coordinates. In this framework the supercharges become the generators of a certain group of translations in superspace. And just as coordinate changes on a real smooth manifold beg the notion of derivatives which are covariant with respect to those changes, superspace naturally motivates \textit{supercovariant derivatives} $D_\rI$ which are covariant with respect to those supertranslations. These derivatives satisfy the same algebra amongst themselves as the supercharges do, namely\footnote{The fact that
the algebras of the $D_\rI $ and $Q_\rI $ operators are identical is a reflection that they simply correspond to the left and right Maurer-Cartan forms on the supermanifold. } 
\begin{equation}
\{  \, D_\rI ~,~ D_\rJ  \, \} ~=~ i \, 2\,   \delta{}_{\rI \, \rJ} \,  \pa{}_{0}  ~~~, 
\end{equation}
and we can specify a supersymmetric system in terms of the action of these derivatives just as well as we can with the supercharges. In the relevant physics literature it is conventional to work largely in terms of the former, and for consistency with cited works we do the same here.

\subsection{The Chiral and Vector Supermultiplets}
\
\\
After a healthy amount of theory and motivation we are ready to examine two concrete realizations of supersymmetry involving fields varying over both space and time \cite{adnkkye}. The first example is the \textit{chiral supermultiplet}: 
\begin{align}
  {\rm D}_a A ~&=~ {\Psi}_a  ~~~,~~~ \cr
{\rm D}_a B ~&=~- \, i \, (\gamma^5){}_a{}^b \, {\Psi}_b  ~~~, \cr
{\rm D}_a {\Psi}_b ~&=~ i\, (\gamma^\mu){}_{a \,b}\,  \partial_\mu A 
~+~  (\gamma^5\gamma^\mu){}_{a \,b} \, \partial_\mu B ~-~ i \, C_{a\, b} 
\,F  ~+~  (\gamma^5){}_{ a \, b} G  ~~, \cr
{\rm D}_a F ~&=~  (\gamma^\mu){}_a{}^b \, \partial_\mu \, {\Psi}_b   ~~~, ~~~ \cr
{\rm D}_a G ~&=~ i \,(\gamma^5\gamma^\mu){}_a{}^b \, \partial_\mu \,  
{\Psi}_b  ~~~.
\label{eq:CM}
\end{align}
 The second is  the \textit{vector supermultiplet}
\begin{align}
{\rm D}_a \, A{}_{\mu} ~&=~  (\gamma_\mu){}_a {}^b \,  {\l}_b  ~~~, \cr
{\rm D}_a {\l}_b ~&=~ - \,i \, \fracm 14 ( [\, \gamma^{\mu}\, , \,  \gamma^{\nu} 
\,]){}_a{}_b \, (\,  \partial_\mu  \, A{}_{\nu}    ~-~  \partial_\nu \, A{}_{\mu}  \, )
~+~ (\gamma^5){}_{a \,b} \,    {\rm d} ~~, {~~~~~~~~~~}  \cr
{\rm D}_a \, {\rm d} ~&=~ i \, (\gamma^5\gamma^\mu){}_a {}^b \, 
\,  \partial_\mu {\l}_b  ~~~.
\label{eq:VM}
\end{align}

A few notes on notation are due at this point.
 $A$, $B$,  $F$ and $G$ are fields with spin $n$ = 0 much like the Higgs boson, $A{}_{\m}$ has $n$ = 2,  $ {\Psi}_a$ and ${\l}_a$ are fields with $n$ = $1/2$,  and $F,G$ and $d$  are \textit{auxiliary fields}, which do not describe physically observable effects
but are nonetheless necessary to maintain supersymmetry.

The $\gamma$ appearing in these equations are \textit{Dirac matrices}, whose defining property is that they form a Clifford algebra with respect to the Minkowski metric. \begin{equation}
\left\{ \gamma^\mu, \gamma^\nu \right\} = 2 \eta^{\mu \nu} \mathbb{I} ~~~.  
\end{equation}
This condition alone does not fix the values of these matrices; we adopt the conventions in \cite{susygenomics} for exact values of these  matrices.

Greek letters - as in the $\mu$ which appears in several of the actions above - range from $0$ to $3$ and are used to index components of tensors over spacetime. The shorthand $\partial_\mu$ refers to the $\mu$-th component of the tuple $(\partial/\partial t, \partial/\partial x, \partial/\partial y, \partial/\partial z) \equiv (\partial / \partial x^0, \partial / \partial x^1, \partial / \partial x^2, \partial / \partial x^3)$.
Latin indices early in the alphabet - $a$, $b$, $c$ above - are so-called spinor indices.
They too indicate components of tensors and here range from 1 to 4.
The term $(\gamma^5)_a{}^b$, for example,  is the  $a,b$ component of the $4 \times 4$ matrix $\gamma^5$.
Both spacetime and spinor indices follow the \textit{Einstein summation convention}, wherein any subscripted index paired with a superscripted matching one is taken to be summed over.
For example, 
\begin{align}
  - i \LP \gamma^5 \RP_a {}^b \Psi_b & \equiv - i \sum _{b = 1} ^4 \LP \gamma^5 \RP_a {}^b \Psi_b ~~~,
\end{align}
\\
\
With notation established we can compute the action of $D_1$ on $B$ in the chiral supermultiplet. Following \cite{susygenomics} we have
\begin{equation}
  \gamma^5
= \begin{pmatrix}
0 & 0 & 0 & i \\
0 & 0 & -i & 0 \\
0 & -i & 0 & 0 \\
i & 0 & 0 & 0
\end{pmatrix} ~~~,
\end{equation}
and  we find
\begin{equation}
{\rm D}_1 B = -i \LP \gamma^5 \RP_1 ^{ \ b} \Psi_b \equiv -i \sum_{b=1}^4  \LP \gamma^5 \RP_1 ^{ \ b} \Psi_b = \Psi_4
~~~.
\label{eq:ex1}
\end{equation}
As discussed earlier,  we can produced a corresponding 1D version of the chiral supermultiplet by formally taking all derivatives $\partial_\mu$ - except for the time-derivative $\partial_0$ - to vanish. For example, the supercovariant derivative of the $F$ field in the chiral multiplet simplifies to 
\begin{equation}
{\rm D}_a F = \LP \gamma^\mu \RP_a ^{ \ b} \partial_\mu \Psi_b \ \to \ \LP \gamma^0 \RP_a ^{ \ b} \partial_0 \Psi_b
~~~.
\end{equation}
\\
\
Again following \cite{susygenomics},  \begin{equation}
  \gamma^0  = \begin{pmatrix}
0 & 1 & 0 & 0 \\
-1 & 0 & 0 & 0 \\
0 & 0 & 0 & -1 \\
0 & 0 & 1 & 0
\end{pmatrix}  ~~~,
\end{equation}
and from this we can read off the transformation laws for $F$ in the reduced version of the chiral multiplet.
\begin{equation}
\begin{cases}
{\rm D}_1 F & = \Psi_2 \\
{\rm D}_2 F & = - \Psi_1 \\
{\rm D}_3 F & = - \Psi_4 \\
{\rm D}_4 F & = \Psi_3
\end{cases} ~~~.
\label{eq:ex2}
\end{equation}

\subsection{Adinkras}
\begin{figure}[htp!]
\centering
\includegraphics[width=0.65\textwidth]{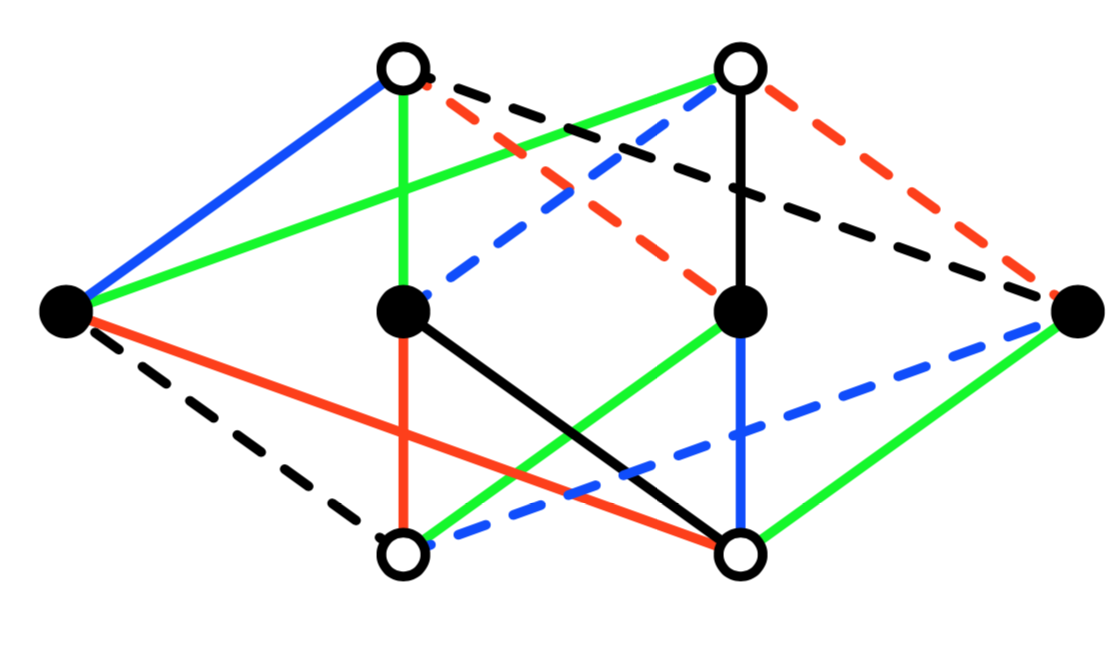}
\includegraphics[width=0.65\textwidth]{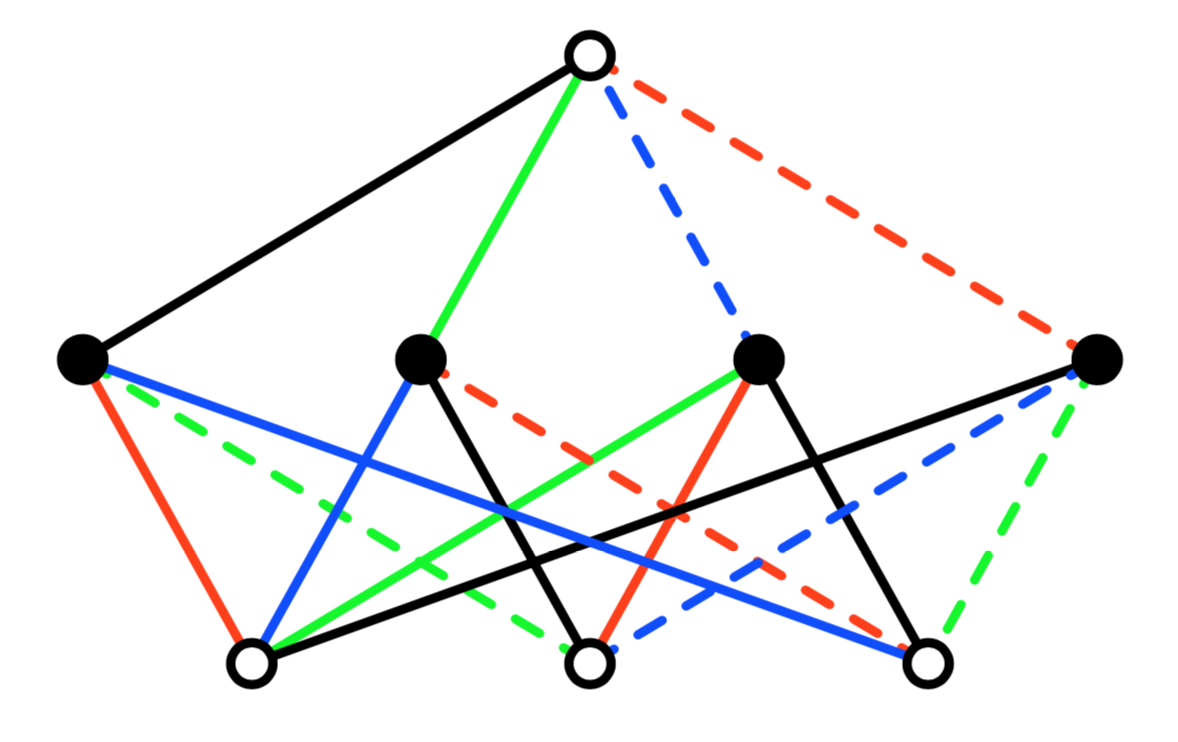}
\caption{The Chiral and Vector Supermultiplets, Dimensionally Reduced and Converted to Adinkras}
\label{Fig:multiplets}
\end{figure}

\
\\
From the transformation laws Eq.\ (\ref{eq:CM}, \ref{eq:VM}) above we see that the supercovariant derivatives in both multiplets take bosonic fields to linear combinations of fermionic fields (sometimes with time derivatives) and vice versa. But the calculations we've seen have taken an especially nice form. Rather then a general linear combination of fermions, each bosonic field has been taken to $\pm$ exactly one fermion Eq.\ (\ref{eq:ex1}, \ref{eq:ex2}). Some further calculations  with these multiplets show that each fermion goes to $\pm$ $ i$ times just one corresponding boson!
With this remarkable property we can build from each multiplet a decorated graph called an \textit{adinkra} by the following procedure. \begin{enumerate}
\item To each field in the multiplet we associate a node in the graph. Visually we represent bosonic and fermionic fields respectively with solid and open circles.
\item Whenever a transformation law takes one field to another, the corresponding nodes are connected by an edge.  An edge coloring indicates which supercovariant derivative was involved. Multiplets like those given here are sometimes called \textit{four-color} systems, and it is conventional to assign to $(\rm D_1, \rm D_2, \rm D_3, \rm D_4)$ some permutation of the colors (red, green, blue, black). An edge is drawn as a dashed line if the corresponding transformation law has a minus sign, and solid otherwise.
\item We arrange the nodes into horizontal rows, with one field appearing in the row directly above another if and only if there exists a transformation law taking the former to the time-derivative of the latter. For systems with corresponding ${\bm \rL}$ and ${\bm \rR}$ matrices, this forces the graph to take the form of a row of fermions sitting above a row of bosons.
\end{enumerate}
Under this procedure, the dimensionally reduced chiral and vector supermultiplets map to the colorful images in Figure~(\ref{Fig:multiplets})
\vskip0.01in
\noindent
\\
\\
\
The Garden algebra realized on such representations then forces a number of regularities onto the graphs.   For example,  every node has exactly one adjacent edge of each color and any length-4 path along alternative colors forms a closed cycle with an odd number of dashed lines. Adinkras map interesting representation-theoretic properties to graph-theoretic ones, \textit{e.g.}, reducible representations correspond to disconnected adinkras. Perhaps most interestingly, Adinkras offer a direct means of {\it constructing} representations of 1D supersymmetry: take a collection of solid nodes, the same number of open nodes, arrange them into alternating rows, then connect them by colored edges with dashings as prescribed by certain combinatorial rules. Since the initiation of this line of thought in \cite{graphtech,graphtheoretic}, the adinkra program has extended from its roots in graph theory and supersymmetry into such languages of Riemann surfaces and Grothendieck ``dessins d' enfants,'' \cite{geometrizationI, geometrizationII}, and error correcting codes 
\cite{errorcorrecting,errorcorrecting2,errorcorrecting3}. It also spurred novel approaches to the \textit{auxiliary field problem} \cite{thinkdifferent}, described further in a later section.

\section{The Adinkra Program - Enter A Coxeter Group}
\
\\
In the work of  \cite{permutadnk}, the real $d \, \times \, d$ matrices that appear in Eq. (\ref{CLFF2}) were explored computationally. The goal of this program was the explicit enumeration and {\textit {construction}} of all linearly 
independent matrices that satisfy the condition in Eq. (\ref{CLFF2}).  The number of matrices found was 384 matrices.  
With a little ``numerology'' one realizes that 384 = 16 $\times$ 24 = $2^4$ $\times$ 4! and this is precisely the order of the Coxeter Group sometimes denoted by the moniker $BC{}_4$.
An explicit examination of all the matrices revealed that each took a form given by
\begin{equation}
 {\bm {\rL}}_{\, \widehat {\rI}} 
 ~=~ 
     {\bm {\cal S}}_{\, \widehat {\rI}}  \, {\bm {\cal P}}_{\, \widehat {\rI}}  ~~~,
\label{aas1}
\end{equation}
where the index $ {\, \widehat {\rI}}$ takes on values 1, 2, $\dots$, 384.  Each of the matrices $  {\bm {\cal S}}_{\, \widehat {\rI}} $,
for fixed value of $ {\, \widehat {\rI}}$, takes the form of a purely diagonal matrix with either plus one or minus one
in each of the diagonal entries. 

Similarly, for each of the matrices $  {\bm {\cal P}}_{\, \widehat {\rI}} $,
for a fixed value of $ {\, \widehat {\rI}}$, takes the form of one of the elements in the permutation group of four
elements.  However, in order to satisfy the condition in Eq. (\ref{CLFF2}), it was found that the entirety of the
permutation group is disaggregated into six distinct {\textit{unordered}} subsets referred to as ``quartets'':\footnote{We use cycle notation in describing permutations.}
\be
\begin{array}{ccc}
~{\bm {\{ {\cal P}{}_{[1]} \}}} & {\bm {\equiv}} &~~ {\bm {\{  \,  (123), \, (134), \, (142) ,  \,  (243)   \,  \} }}  ~~~,  \\
{~} & {~} &  \\
~{\bm {\{  {\cal P}{}_{[2]}  \}}} & {\bm {\equiv}} &~~ {\bm {\{  \,  (124), \, (132), \, (143) \, , (234)    \,  \} }}  ~~~,  \\
{~} & {~} &   \\
~{\bm {\{  {\cal P}{}_{[3]}  \}}} & {\bm {\equiv}} &~~ {\bm {\{  \, (14),  \,  (23),  \, (1243) , \, (1342)    \,  \} }}  ~~~,  \\
{~} & {~} &    \\
~{\bm {\{  {\cal P}{}_{[4]}  \}}} & {\bm {\equiv}} &~~  {\bm {\{   (13), \, (24),   \, (1234), \,   \, (1432)  \,  \} }}  ~~~,  \\
{~} & {~} &    \\
~{\bm {\{  {\cal P}{}_{[5]}  \}}} & {\bm {\equiv}} &~~ {\bm {\{ \, (12), \, (34),  \, (1324) , \,  (1423)     \,  \} }}   ~~~,  \\
{~} & {~} &   \\
~{\bm {\{  {\cal P}{}_{[6]}  \}}} & {\bm {\equiv}} &~~ {\bm {\{    \,  (), \, (12)(34) ,   \,  (13)(24) , \,  (14)(23) \,   \} }}
~~~.
\end{array}
\label{KP}
\ee

Stated another way, if one makes an arbitrary choice among the matrices in Eq.\ (\ref{aas1}) and considers { \textit {only}}
its dependence on the permutation matrix that appears, the remaining three permutation elements  must be in the same subset that is shown in Eq. (\ref{KP}) in order for the collection to
satisfy Eq.\ (\ref{CLFF2}).

There was one other property noted in \cite{permutadnk} about this disaggregation. 
The concept of Hodge Duality of forms is  well established in differential geometry.
Essentially, an operation was uncovered on the 
collection of disaggregated quartets that ``shadows'' the four dimensional Hodge Duality acting on the supermultiplets
of field variables.
\begin{equation}
 \begin{array}{r@{\>=\>}l}
  {}^{\bm*}\{ {\bm {\cal P}{}_{[1]}}  \}&\{ {\bm {\cal P}{}_{[2]}}  \}~,\\
  {}^{\bm*}\{{\bm {\cal P}{}_{[2]}}  \}&\{ {\bm {\cal P}{}_{[1]}}  \}~,  \\
 \end{array}
 \qquad
 \begin{array}{r@{\>=\>}l}
  {}^{\bm*}\{{\bm {\cal P}{}_{[3]}}  \}  &\{ {\bm {\cal P}{}_{[3]}}  \}~,\\
  {}^{\bm*}\{{\bm {\cal P}{}_{[4]}}  \}&\{ {\bm {\cal P}{}_{[4]}}   \}~,\\
  {}^{\bm*}\{ {\bm {\cal P}{}_{[5]}}\}& {\bm \{{\cal P}{}_{[5]}}  \}~,\\
 \end{array}
 \qquad
  {}^{\bm*}\{{\bm {\cal P}}{}_{[6]} \}=\{ {\bm {\cal P}}{}_{[6]} \}~.
  \label{Hodge}
\end{equation}
The way the ``star operation'' is realized is very direct algebraically.
When permutation elements are written in terms of cycles, the standard convention is to
read them from left to right.  So for example the symbol $(234)$ implies a permutation where  1 $\to$ 1, 
2 $\to$ 3,  3 $\to$ 4, and 4  $\to$ 2.
Under the star operation, one reads instead from right to left.  Therefore, $
{}^{\bm*}(234)$ implies a permutation where 1 $\to$ 1, 2 $\to$ 4, 4
$\to$ 3, and 3 $\to$ 2. The results shown in Eq.\ (\ref{Hodge}) depend
critically on the fact that the subsets $\{ {\bm {\cal P}{}_{[1]}}
\}$, $\dots$, $\{ {\bm {\cal P}{}_{[6]}} \}$ are {\textit{unordered}} sets.
 
A final observation is an analogy to genomics.  Projecting down from the systems Eq.\ (\ref{eq:CM}) and 
Eq.\ (\ref{eq:VM})
leads to the subsets $\{ {\bm {\cal P}{}_{[1]}} \}$, $\dots$, $\{ {\bm {\cal P}{}_{[3]}} \}$.
The information contained in these sets of ``superdifferential'' equations has been transferred to the domain of
 a Coxeter group and the permutation cycle structure of its elements.  It is not too terrible an abuse of
language to say the process ``sequences'' the initial equation to reveal a genomic-like substructure.

\section{The Optimization Program - Supersymmetry from Scratch}
\
\\
We close with a high-level discussion of a more recent approach to generating representations of supersymmetry, for which we return to the earlier convention of working in terms of supercharges rather than supercovariant derivatives.

\subsection{Auxiliary Fields}
\
\\
One of the first physical results of supersymmetry is that it demands that a theory contain an equal number of bosons and fermions. This implies the more general fact that many theories contain obstructions - sometimes as simple as this particle number mismatch, sometimes more subtle - to the existence of supercharges satisfying the constraints of supersymmetry. The \textit{auxiliary field problem} asks whether a given physical theory can be augmented by the addition of new bosons and fermions (the titular auxiliary fields) in order to allow for supersymmetry. \\
\\
\
In general this problem interacts in difficult ways with the dynamical laws of motion of a system, which we have completely neglected in the present discussion. In the toy model of supersymmetric quantum mechanics however it becomes slightly more tractable, and in some sense reduces to the problem of taking a fixed collection of matrices and padding them with appropriate entries so that they satisfy the Garden Algebra Eq. (\ref{gardenalgebra}). This computational approach
was {\it {initiated}} in \cite{thinkdifferent}.  
Essentially, this work proposed an {\it {ab}} {\it {initio}} approach where the 2$d^2$ entries of the ${\bm {\rL}}_\rI$ and ${\bm {\rR}}_\rI$ matrices be regarded as a set of coordinates describing a  2$d^2$-dimensional manifold.
With this interpretation, the constraints in Eq.\ (\ref{CLFF2}) on these coordinates describe a real algebraic variety in
terms of polynomials on these coordinates.  This high dimensional surface may be called ``the adinkra continent of SUSY representations'' as the zeros of this surface define supersymmetric representations in terms of ${\bm {\rL}}_\rI$ and ${\bm {\rR}}_\rI$ matrices.  Using examples and simplifying assumptions about some matrix entries, in \cite{thinkdifferent} the 
{\it {first}} demonstration of practical
computational algorithm design showed the feasibility of finding solutions.

\subsection{Supersymmetry as Optimization}
\
\\
Expanding this approach to consider the edge case in which we take all of the matrix entries to be unknown. That is, we fix $\mathcal{N}$ and $d$ and search the space of all tuples of $d \times d$ real matrices $\LP {\bm \rL}_1\RP$ for solutions to Eq. (\ref{gardenalgebra}). Since Eq. (\ref{gardenalgebra}) is a system of constraints which are polynomial in the components of those matrices, this reduces to doing real algebraic geometry over $\R^{2\mathcal{N}d^2}$ with varieties with large amounts of both discrete and continuous symmetry. It also suggests the more analytic approach of converting this system of constraints into a single cost function, it turns out that the global minima are in bijection with realizations of the Garden algebra.  Both of these lines of thought are explored in a coming work.

\vspace{.05in}
 \begin{center}
\parbox{4in}{{\it ``Sometimes it is the people no one can imagine anything
 \\ $~~$
of who do the things no one can imagine.'' \\ ${~}$
\\ ${~}$ }\,\,-\,\, Alan Turing}
 \parbox{4in}{
 $~~$} 
 \end{center}
 \noindent
{\bf {Acknowledgments}}\\[.1in] \indent

The research of S.\ J.\ G.\ is supported in part by the endowment of the Ford Foundation 
Professorship of Physics at Brown University and also grateful acknowledgment of support 
from the Brown Theoretical Physics Center is given.


\end{document}